\documentclass[nofootinbib,superscriptaddress,twocolumn,aps,prc,showpacs,10pt]{revtex4-1}

\usepackage{amsmath}
\usepackage{amssymb}
\usepackage{graphicx}
\usepackage{mathrsfs}
\usepackage{color}
\newcommand{\bvec}[1]{\mbox{\boldmath $#1$}}

\begin{document}
\title{Collective aspects deduced 
from time-dependent microscopic mean-field with pairing: application to the fission process}

\author{Yusuke Tanimura} \email{tanimura@ipno.in2p3.fr}
\author{Denis Lacroix} \email{lacroix@ipno.in2p3.fr}
\affiliation{Institut de Physique Nucl\'eaire, IN2P3-CNRS, Universit\'e Paris-Sud, F-91406 Orsay Cedex, France}
\author{Guillaume Scamps} 
\email{scamps@nucl.phys.tohoku.ac.jp}
\affiliation{ Department of Physics, Tohoku University, 
Sendai 980-8578, Japan}
\date{\today}

\begin{abstract}
Given a set of collective variables, a method is proposed to obtain the associated conjugated collective momenta and masses 
starting from a microscopic  time-dependent mean-field theory.  The construction of pairs of conjugated variables is the first step to
bridge microscopic and macroscopic approaches. The method is versatile and can be applied to study a large class of nuclear 
processes. An illustration is given here with the fission of $^{258}$Fm. Using the quadrupole moment and eventually higher-order 
multipole moments, the associated collective masses are estimated along the microscopic mean-field evolution. When more 
than one collective variable are considered, it is shown that the off-diagonal matrix elements of the inertia play a crucial role. Using the information 
on the quadrupole moment and associated momentum, the collective evolution is studied. It is shown that dynamical effects beyond the adiabatic limit
are important. Nuclei formed after fission tend to stick together for longer time leading to a dynamical scission point at larger distance between nuclei compared to the one anticipated from the adiabatic energy landscape. 
The effective nucleus-nucleus potential felt by the emitted nuclei is finally extracted. 
\end{abstract}


\maketitle

\section{Introduction}

Nuclear time-dependent mean-field based on the energy density 
functional approach is experiencing nowadays a renewal of interest
\cite{Kim97,Sim01,Nak05,Mar05,Uma05,Was08}.
In particular, it allows one to describe a wide variety of dynamical 
processes ranging from small to large amplitude collective motions, including 
nuclear reactions. Among the most difficult challenges, one can mention 
the description of nuclear fission where a single 
nucleus encounters large deformation leading finally to separated fragments. 

Although the fission process is quite well understood phenomenologically \cite{Van73}, it still
 remains one of the most difficult process to describe microscopically. One of the major difficulty 
is the necessity to treat collective and single-particle degrees of freedom (DOF) simultaneously
as quantum objects \cite{Neg89}. 
Moreover, fission is a dynamical process and therefore 
should be treated as such. 
To describe the dynamic of fission, one can a priori
use two strategies. Using the fact that the time-scale associated to fission 
is rather large, the most common starting point is to first select a few collective 
DOFs and generate an adiabatic energy landscape. Then, the time-scale
associated to fission can be evaluated using semi-classical approximation. Alternatively, 
the real time dynamic can be explicitly followed using for instance the 
Time-dependent Generator coordinates 
method \cite{Gou05}. Two important problems are generally encountered in this strategy. 
First, this approach 
can hardly accommodate too many collective DOFs due to the increasing complexity. Second, while rather 
slow, the dynamic of fission might deviate from the completely adiabatic path 
when the two fragments approach the scission.

An alternative strategy is to use microscopic quantum transport theories such as the time-dependent energy density 
functional (TD-EDF) theory. This
approach offers the possibility to describe 
some aspects of the fission process without assuming adiabaticity while leaving the possibility 
to explore rather complex shapes during the separation process. In addition, the recent inclusion 
of dynamical pairing has opened new perspectives \cite{Ave08,Eba10,Ste11,Sca12,Has12,Sca13}. 
On the other hand, the TD-EDF approach cannot 
describe completely the fission process due to the absence of spontaneous symmetry breaking and due to 
poor treatment of quantal effects in collective space. However, it can still provide important information in particular 
after the system has passed the fission barrier. This has been illustrated in Refs. \cite{Sim14,God14,God15} 
and more recently in Ref. \cite{Sca15}.  
 
The TD-EDF directly performs the evolution of single-particle states in a self-consistent mean-field. From this evolution, one can 
directly infer the information on any one-body degree of freedom like multipole deformation, neck formation and/or fragment 
separation, final kinetic energies ... The aim of the present work is to explore the possibility to get macroscopic 
transport coefficients, like collective mass, collective potential or energy dissipation, directly from TD-EDF.  In particular, this should 
allow us to compare the result of TD-EDF to similar quantities generally obtained in the adiabatic limit and/or macroscopic models.    

In the following, we analyze first how collective masses and momenta can be associated to given collective observables along a 
microscopic mean-field path. Once the pairs of conjugated operators are available, a macroscopic reduction of the microscopic 
approach can be made to give physical insights. 
For the fission process, the energy sharing between internal and selected DOFs can 
be precisely scrutinized.   

\section{Collective mass and momentum extracted from dynamical mean-field theory}\label{sec:mass}

In the present section, we assume that the mean-field trajectory including or not pairing 
is known, leading to a specific trajectory in the Liouville space of the normal and anomalous density matrices
 $(\rho(t), \kappa(t))$. Starting from these densities, we want to extract 
information on a set of given collective variables. Note that, although we present examples 
specifically on the fission process, the approach developed here is general and can be applied 
to other processes.    

Let us consider a given collective DOF associated to the one-body operator $\hat Q_\alpha$. 
We restrict the present discussion to the case where  $\hat Q_\alpha$
 corresponds to a local one-body operator, i.e.
 \begin{eqnarray}
\hat Q_\alpha = \int d^3r~ Q_\alpha (\bvec r) \hat \Psi^{\dagger} (\bvec r) \hat \Psi (\bvec r) ,
\end{eqnarray}
where $\hat \Psi^{\dagger} (\bvec r)$ and $ \hat \Psi (\bvec r) $ are the creation/annihilation operators of a particle at position 
$\bvec r$. For the sake of simplicity, we omitted the spin and isospin quantum numbers. Note that most macroscopic 
DOFs of interest like multipole operators, relative distance, mass asymmetry, ... correspond to expectation values 
of local one-body operators. 

Along the mean-field trajectory, the collective evolution is given by:
\begin{eqnarray}
q_\alpha (t) & \equiv & {\rm Tr}\left( Q_\alpha \rho (t) \right) = \int Q_\alpha (\bvec r ) n(\bvec r,t ) d^3r,
\end{eqnarray} 
where we have introduced the local density $n(\bvec r,t) = \langle \bvec r | \rho(t)| \bvec r \rangle $. 
The first step to bridge the microscopic mean-field theory and a macroscopic like evolution for the 
collective variable $q_\alpha$ is to find the corresponding conjugated momentum $p_\alpha$ and associated 
collective mass, denoted by $M_\alpha$. Assuming a classical equation of motion for $q_\alpha$, one should 
fulfill the simple constraint $\dot q_\alpha = p_\alpha/M_\alpha$. In the TD-EDF approach, we have:
\begin{eqnarray}
i\hbar \frac{d \langle Q_\alpha \rangle}{dt} & = &  {\rm Tr} \left( Q_\alpha [h(\rho), \rho]  \right) , \nonumber 
\end{eqnarray}
where $h(\rho)$ is the time-dependent mean-field hamiltonian.
Using properties of the trace and assuming a local mean-field potential, we obtain:
\begin{eqnarray}
\frac{d \langle Q_\alpha \rangle}{dt} & = &  -\frac{i}{2\hbar m} {\rm Tr} \left( \left[ Q_\alpha ,  p^2\right] \rho (t)  \right) , 
\label{eq:qdot}
\end{eqnarray}
where $m$ is the nucleon mass. 
From this expression, a simple guess for the conjugated momentum is to assume that it can directly be defined 
through:
\begin{eqnarray}
\hat P_\alpha & \equiv & -i \frac{M_\alpha}{2\hbar m} \sum_{ij} \langle i | [ Q_\alpha , p^2] |j\rangle  a^\dagger_i a_j,
\end{eqnarray}   
where $(a^\dagger_i, a_i)$ are creation/operators of any complete single-particle basis. From this definition, 
one can easily see that the matrix elements of the operator $\hat P_\alpha$ are given by:
\begin{eqnarray}
P_\alpha & \equiv & - i \hbar \frac{M_\alpha}{m} \left( \frac{\nabla^2 Q_\alpha }{2} +\nabla Q_\alpha . \nabla \right) , 
\label{eq:momentum}
\end{eqnarray}  
that is similar to the expression obtained in \cite{Ber97} using a variational principle around a static mean-field 
to study anharmonic effects in giant resonances.

One shortcoming of the above expression is that the operator $\hat P_\alpha$ contains the collective mass $M_\alpha$ that is 
unknown. To further progress, we  seek for the condition that $(\hat Q_\alpha , \hat P_\alpha)$ are conjugated observables. 
Similarly to the Time-Dependent RPA (TDRPA), we need to impose the condition:
\begin{eqnarray}
{\rm Tr}\left( \rho(t) [\hat Q_\alpha, \hat P_\alpha ]\right)= i\hbar \label{eq:conj} ,
\end{eqnarray}
along the trajectory. We have: 
\begin{eqnarray}
\langle  [  \hat Q_\alpha , \hat P_\alpha ] \rangle &=& {\rm Tr} \left( [Q_\alpha, P_\alpha] \rho (t) \right) = +i \hbar \frac{M_\alpha}{m} \langle  \nabla Q_\alpha . \nabla Q_\alpha \rangle\nonumber
\end{eqnarray} 
We see that the condition  (\ref{eq:conj}) determines uniquely the collective mass through: 
\begin{eqnarray}
\frac{1}{M_\alpha (t)} &=& \frac{1}{m} {\rm Tr} \left[ \rho(t) \nabla Q_\alpha . \nabla Q_\alpha  \right] \label{eq:mass1}, 
\end{eqnarray}
all along the trajectory and henceforth also leads to an unambiguous 
definition of the collective momentum $\hat P_\alpha$ when reporting the mass 
in Eq. (\ref{eq:momentum}). 
Similar formula is sometimes used to compute collective mass from a microscopic adiabatic 
energy landscape (see for instance \cite{Rei90,Rei92,Ada95}). 
The difference is that this expression has been derived here without assuming adiabaticity. In addition, 
since the expectation value is directly performed using the time-dependent mean-field density it automatically 
contains possible 
influence of other DOFs as well as the pairing effects.  

The mass formula (\ref{eq:mass1}) is rather straightforward to calculate. We give illustration of some expression obtained for specific collective operators in Appendix \ref{sec:appendix}.

Once the mass and the momentum are known, one can also define the collective kinetic energy 
corresponding to the selected variable as 
\begin{eqnarray}
E_{\rm kin}^\alpha (t) = \frac{p_\alpha^2 (t)}{2M_\alpha}=\frac{1}{2}M_\alpha \dot q_\alpha^2 (t). 
\end{eqnarray}

\subsection{Generalization to several collective degrees of freedom.}
\label{sec:diagm}

Let us now consider a more general case where a set of $N$ collective DOFs $\{ Q_\alpha \}_{\alpha=1,N}$ are selected. 
A naive generalization to previous section is to assign to each variable $Q_\alpha$, a collective momentum $P_\alpha$
with matrix elements given by Eq. (\ref{eq:momentum}). One should a priori also 
generalize the commutation relation (\ref{eq:conj}). Using Eq. (\ref{eq:momentum}), we have:\begin{eqnarray}
\langle[Q_\alpha,P_\beta]\rangle&=&
i\hbar\frac{M_{\beta\beta}}{m}\langle \nabla Q_\alpha\cdot\nabla Q_\beta\rangle =
i\hbar\frac{M_{\beta\beta}}{M_{\alpha\beta}}, 
\end{eqnarray}
where the off-diagonal mass matrix elements reads
\begin{eqnarray}
\frac{1}{M_{\alpha \beta} (t)}&=& 
\frac{1}{m} {\rm Tr} \left[  \rho(t) \nabla Q_\alpha . \nabla Q_\beta \right]. \label{eq:mass2}
\end{eqnarray}
This expression naturally extend the previous case and was also given in Ref. \cite{Ada95}. 
As shown in Appendix \ref{sec:mdiag}, the diagonalization of the mass gives new canonical pairs of 
collective operators $(\hat Q'_k, \hat P'_k)$, whose commutation rules identifies with the TDRPA ones and 
are given by
\begin{eqnarray}
\langle  [\hat Q'_k, \hat P'_l ] \rangle = i\hbar \delta_{kl} . \label{eq:conjstrong} 
\end{eqnarray}
The diagonalization is equivalent also to removing the correlation among the variables 
like multipole moments. 

These new operators are particularly useful to get simple expressions for the evolution and collective energy. 
In particular, we have $\dot Q'_\alpha = P'_\alpha/M'_\alpha$ while the collective kinetic energy is simply given by
\begin{eqnarray}
E_{\rm kin}^{\{\alpha\}}=\sum_k \frac{{P'}^2_k(t) }{2 M'_{k} (t)}. \label{eq:ekingen}
\end{eqnarray}

Once the set of collective variables is properly defined, macroscopic analysis of TD-EDF evolution 
can be made. Such a connection from the microscopic level to the macroscopic one is illustrated below for the fission process.

\section{Application to the fission of $^{258}$Fm}

To illustrate the method presented in the previous section, we consider the case of 
$^{258}$Fm that was the subject of the recent work \cite{Sca15}. This nucleus is anticipated 
to have three different paths towards fission. In this work, we concentrate on the so-called symmetric compact shape. 
The energy landscape is obtained using the EV8 program with a constraint of the quadrupole 
moment \cite{Bon05}. We use here the standard definition for multipole moments
\begin{equation}
Q_\lambda= \sqrt{\frac{16\pi}{2\lambda+1}}\langle r^\lambda Y_{\lambda 0} \rangle,
\end{equation}
leading for instance to $Q_2(\bvec r)=2z^2-x^2-y^2$.   

\begin{figure}
\begin{center}
\includegraphics[scale=0.7]{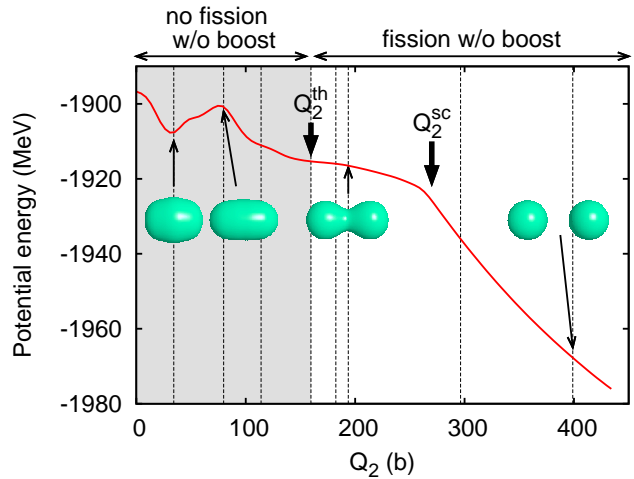}
\end{center}
\caption{(Color online) Potential energy curve of $^{258}$Fm nucleus as a function of the 
quadrupole deformation parameter (in barn unit). 
Isosurfaces of the total density drawn at half the maximum value at $Q_2=34$ b, 80 b, 194 b, and 399 b 
are also shown. The horizontal lines indicate the different starting points that are used in this work 
as initial conditions for the time-dependent evolution. 
The different vertical dashed lines corresponds from left to right to
$Q_2=34.2$ b, $Q_2=80$ b (barrier position), $Q_2=160$ b (spontaneous fission threshold $Q^{\rm th}_2$) 
, $Q_2=182$ b, $Q_2=194$ b, $Q_2=296$ b, and $Q_2=400$ b. 
The two thick arrows indicate the spontaneous fission threshold $Q_2^{\rm th}$ and the adiabatic scission 
point $Q_2^{\rm sc}$.}
\label{fig:pes}
\end{figure}

An illustration of the potential energy curve (PEC) is shown in Fig. \ref{fig:pes} 
as a function of the quadrupole moment $Q_2$. As in ref. \cite{Sca15}, the Sly4d 
Skyrme functional \cite{Kim97} is used for the mean-field channel while a constant interaction 
is retained for the pairing channel. The static calculations are performed with a mesh size 
$13.2 \times 24.4 \times 13.2$ fm$^3$
and a mesh step  $\Delta x=0.8$ fm.

The dynamical evolution of the system starting from any point of the PEC can be made consistently 
using the recently developed TD-EDF code including pairing in the BCS approximation \cite{Sca13,Sca13b,Sca14}. 
Dynamical calculations shown here are performed in a mesh of size $26.4\times 72.8 \times 13.2$ fm$^3$ with 
the same mesh step as in the static case $\Delta x=0.8$ fm. 
The time step is $\Delta t=1.5\times 10^{-24}$ sec $\approx 0.45$ fm/c. 
In the present calculations, reflection and axial 
symmetries are assumed in the constrained calculation. Since symmetry cannot be broken spontaneously 
by mean-field, only even multipole moments can be non-zero during the evolution. In particular, we do not consider 
here possible octupole deformation.   

\begin{figure}
\begin{center}
\includegraphics[scale=0.7,angle=0]{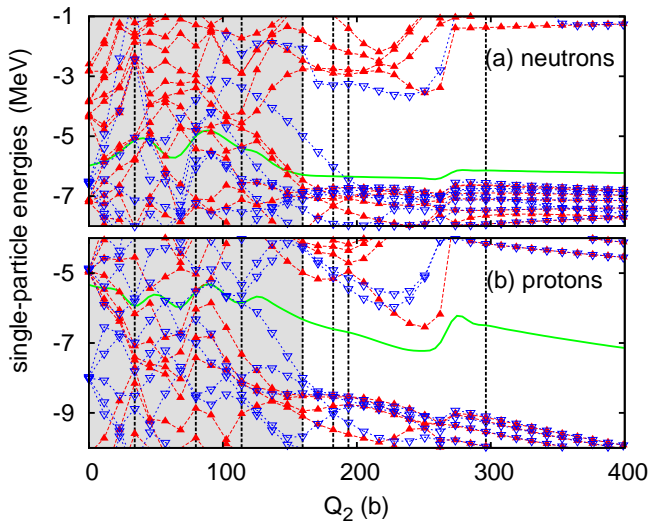}
\end{center}
\caption{(Color online) Single-particle energies in $^{258}$Fm nucleus along the adiabatic PEC. 
The green solid curves show the neutron and proton Fermi energies. 
Positive and negative parity states are respectively shown with red filled and blue open triangles. 
The vertical lines indicate the initial values of the quadrupole moment taken in the present dynamical calculations. The shaded area presenting the region where the system does not spontaneously fission is also shown.}
\label{fig:esp}
\end{figure}

As it was observed previously including or not pairing, the system will spontaneously separate into two fragments   
only above a certain value of the initial quadrupole moment, which is larger than that of the fission barrier shown in Fig. \ref {fig:pes} \cite{Sim14, God14, God15, Sca15}. 
The lowest initial quadrupole moment leading to spontaneous fission within TD-EDF is called hereafter ``Dynamical Fission Threshold'' and 
will be denoted by $Q^{\rm th}_2$. In the present calculation, the threshold deformation is approximately 
$Q^{\rm th}_2 \simeq 160$ b.
The shaded area in this figure indicates the region where the system does not spontaneously fission. 
The fact that $Q^{\rm th}_2$ is well beyond the expected barrier position signs the deviation from 
the adiabatic limit of the microscopic transport theory close to single-particle levels crossing. This point was 
already discussed in Ref. \cite{Sim14}. To illustrate the connection between the dynamical fission threshold and level crossing, 
the single-particle energies evolution obtained in the static constrained mean-field are 
shown in Fig. \ref{fig:esp} as a function of $Q_2$.
 
We see in particular that for large $Q_2$ a gap in single-particle energies appears that signs the transition from one to two
nuclei.
At low quadrupole moments, many crossing occurs. When one leaves the system initially in the shaded area, single-particle wave functions will evolve in time. 
However, the motion is not adiabatic and both occupations above and below the Fermi energy will be populated in time. The PEC is meaningful only if lowest levels are preferentially occupied during the evolution while higher levels are depopulated.  TD-EDF including
or not pairing does not lead to sufficient reorganization of single-particle occupation numbers during 
the crossing to follow the adiabatic PEC as it has already been realized in Refs. \cite{Sim14,God14,God15,Sca15}. 
As studied in Ref. \cite{God14}, the system initialized inside the shaded area can still fission 
if, for instance, a boost in the quadrupole momentum is applied at initial time. 
In practice, the boost is imposed by applying the local operator 
$\displaystyle \exp(i p_2 Q_2(\bvec r)/ \hbar)$ to each single-particle wave-function. This induces an additional 
initial collective kinetic energy \cite{God14} :
\begin{eqnarray}
E^{\rm ini}_2 &=& \frac{p_2^2}{2m} \int  |\nabla Q_2(\bvec r)|^2 n(\bvec r, t=0)d^3r \nonumber
\end{eqnarray}
where $n(\bvec r, t=0)$ denotes the local density of the system in the adiabatic curve selected at a given initial moment. 
Note that $E_2^{\rm ini}=p_2^2/2M_2$ with the quadrupole mass given by Eq. (\ref{eq:mass1}). 

\begin{figure*} [htbp]
\begin{center}
\includegraphics[scale=2.2]{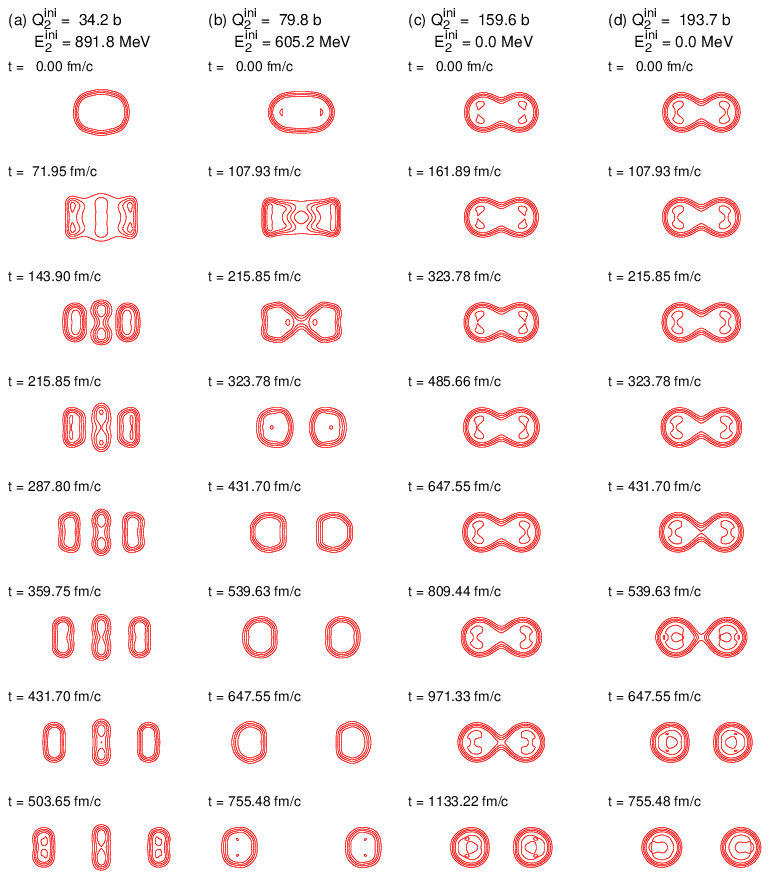}
\caption{(Color online) Density profiles obtained for different initial $Q_2$. The densities are shown as a	 function 
of time. For initial $Q_2$ below the fission threshold, a quadrupole boost has been imposed initially. From left to right, the initial
$Q_2$ values are  $Q_2=34.2$ b, $Q_2=80$ b, $Q_2=160$ b and $Q_2=194$ b. The system is eventually initially boosted leading 
to non zero values of $E^{\rm ini}_2$ directly indicated in the figure. The iso-density curves are drawn from 0.03 to 0.15 fm$^{-3}$ with increment of 
0.03 fm$^{-3}$. }
\label{fig:denstime}
\end{center}
\end{figure*}

In Fig. \ref{fig:denstime},  several examples of density evolution obtained for different initial conditions including or not 
a boost initially and leading to fission are illustrated. 
We see that a variety of phenomena including ternary fission in some cases can be observed. 

\subsection{Mass parameter from TD-EDF}

In the present section, we consider different initial quadrupole deformations between the fission barrier and the scission point. The scission point corresponding to a quadrupole deformation $Q^{\rm sc}_2$ can already be seen in the
shown in Fig. \ref{fig:pes}. It corresponds to the kink in the PEC appearing at $Q_2 \simeq 270$ b. 
After the scission point, the PEC is nearly dominated by the Coulomb interaction between the two fragments (See also Fig. 
\ref{fig:force}). 

As an illustration, we consider that the initial state corresponds to $Q^{\rm ini}_2 = 160$ b, that is a situation just above 
the spontaneous fission threshold. This initial condition is similar to the one considered in Ref. \cite{Sca15}. In particular, it has been shown that if the system is left initially with zero collective energy, the total final kinetic energy  of fragments after TD-EDF evolution 
is compatible with experimental observation. To study possible non-adiabatic effect, initial conditions with boost of varying 
intensity (including no boost at all) are used. 
\begin{figure}
\begin{center}
\includegraphics[scale =.9]{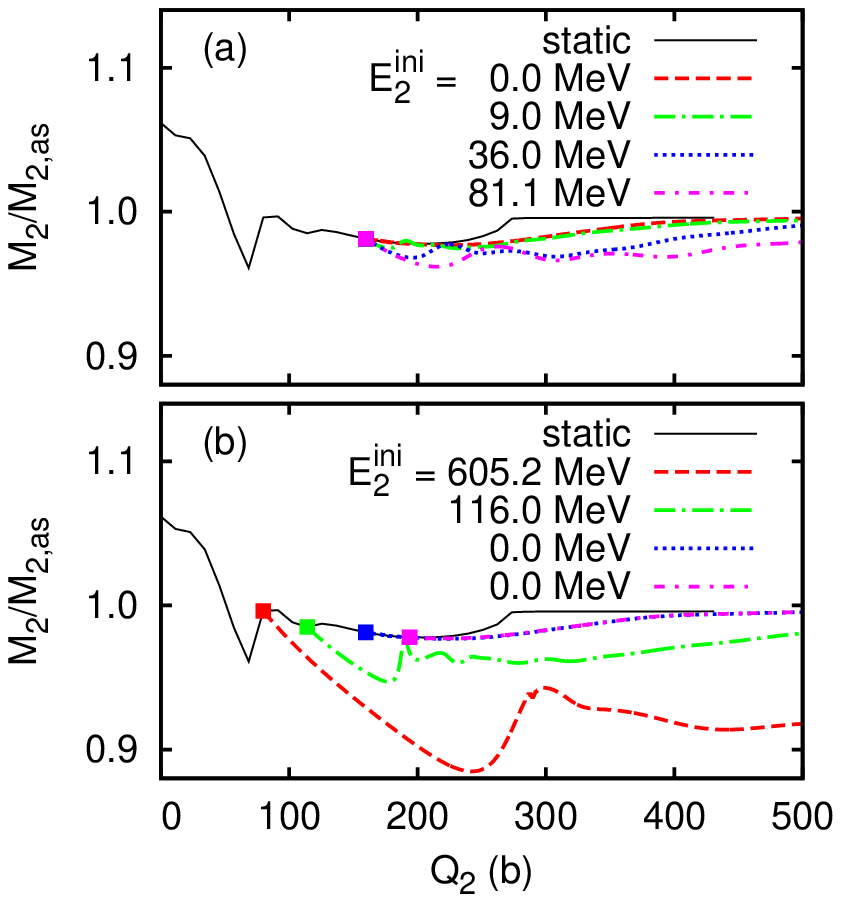}
\end{center}
\caption{(Color online) Top: Quadrupole mass parameter calculated from 
TD-EDF paths. In all cases, the  initial quadrupole moment is $Q^{\rm ini}_2=160$ b.
Different trajectories correspond to different initial boosts. The corresponding initial collective energies $E_2^{\rm ini}$ 
are systematically reported in the figure. 
The quadrupole mass obtained using Eq. (\ref{eq:mass1}) assuming that the system follow the adiabatic PEC 
is also shown (solid line) for comparison. 
Bottom: Quadrupole mass obtained with varying initial quadrupole 
deformation, $Q^{\rm ini}_2=80$ b, 114 b, 160 b, and 194 b. }
\label{fig:mass_b14}
\end{figure}

The quadrupole moment is selected as the most relevant collective DOF. Using Eq. (\ref{eq:mass1}), the associated collective mass 
is then given by (appendix \ref{sec:appendix})
\begin{eqnarray}
\frac{1}{M_\alpha}= \frac{4}{m} \left(2 \langle r^2(t) \rangle+ \langle Q_2 (t) \rangle \right), 
\end{eqnarray}
where $\langle r^2(t) \rangle$ and $\langle Q_2 (t) \rangle$ are respectively the root-mean-square radius and quadrupole moment along the path.  

To get physical insight it is interesting to consider the situation where the system is already 
about to get separated into two fragments with a neck. 
Assuming simply that the neck position is at the center of the whole system, quantities like 
mass, position, momentum, and intrinsic deformations of each fragment can be estimated through:
\begin{eqnarray}
\langle X (t) \rangle_{[1]}  & = & \int X(\bvec r) n(\bvec r) \Theta(z) d^3 \bvec r , \nonumber \\
\langle X (t) \rangle_{[2]}  & = & \int X(\bvec r) n(\bvec r) \left[ 1 - \Theta(z)\right] d^3 \bvec r , \nonumber 
\end{eqnarray}
where $X$ is the local operator corresponding to the specific quantity under interest, $\Theta(z)$ is the Heaviside step 
function and $[i=1,2] $ is a label of fragments. For a di-nuclear system, the quadrupole mass can be recast as:
\begin{eqnarray}
\frac{1}{M_2}&=& \frac{8 \mu(t)}{m^2} R^2(t) + \frac{4}{m} \sum_{i=1,2}\left[ 2 \langle r^2 (t) \rangle_{[i]} + \langle Q_2 (t) \rangle_{[i]} \right]
\nonumber
\end{eqnarray}
where $\mu(t) = m A_1 A_2 /A$ is the reduce mass of the system and $R(t)$ is the relative distance between the center of mass of the two fragments.

At very large distance, $R(t) \rightarrow + \infty$, we see that the mass is dominated by the first term and tends to infinity.  
For display purpose we consider, as a reference mass, the mass obtained assuming no intrinsic quadrupole deformation 
$\langle Q_2 (t) \rangle_{[i=1,2]} =0$ and using the simple prescription  $\langle r^2 (t) \rangle_{[i]} = \frac{3}{5}  A_i \left(1.2 A^{1/3}_i \right)^2$ fm$^{2}$. The reference mass obtained in this way is denoted by $M_{2,{\rm as}}$. In the following, the quadrupole mass will always be shown with respect to this mass. 

In Fig. \ref{fig:mass_b14}-a, the ratio $M_2/M_{2,{\rm as}}$ of quadrupole mass deduced 
with the present method is shown as a function of $Q_2$ along the TD-EDF trajectories for
 $Q^{\rm ini}_2=160$ b. In this figure, the mass obtained using increasing initial boost 
in the quadrupole collective  momentum are shown. 
To illustrate the departure from the adiabatic path, we also show the result 
 obtained for a given $Q_2$ assuming that the local density identifies with the corresponding 
 density  directly obtained from the constrained mean-field calculation. In the following, the latter is referred to as 
 ``static mass''. 
 
 We observe in Fig. \ref{fig:mass_b14}-a that the mass is in general rather close to the 
 static mass, especially if $E_2^{\rm ini}=0$ MeV. In that case, the system first 
 follows closely the adiabatic case and then some deviation is observed. The deviation occurs around the scission point. 
At this point, the slope of the PEC suddenly change to match with the Coulomb case that dominates at large distance. 
This increase of slope is expected to induce also a larger collective velocity and therefore also induce a possible departure 
from the adiabatic limit. We see in this figure that the mass also depends on the initial collective velocity imposed 
to the system. The larger is the initial velocity, the more deviation from the static mass is observed. 
It is however worth mentioning that the adiabatic/non-adiabatic behavior cannot easily be concluded solely from 
the difference of mass as will be further illustrated below.

\subsection{Mass parameter for $Q^{\rm ini}_2 \le  Q^{\rm th}_2$}

\begin{figure}
\begin{center}
\includegraphics[scale = .6 ,angle = 0]{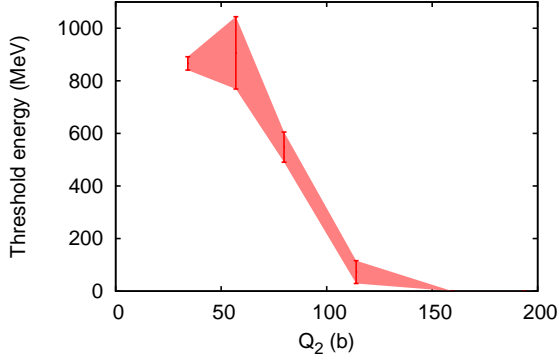}
\end{center}
\caption{Minimum collective energy after a boost that should be initially deposited 
into the system to induce the fission of $^{258}$Fm for $Q^{\rm ini}_2 \le Q^{\rm th}_2$. 
This energy is plotted as a function of the initial quadrupole deformation considered. 
For each $Q_2$, the highest (resp. lowest) value of the initial collective energy where the scission is not 
(resp. is) observed is reported.}
\label{fig:eth}
\end{figure}

As we have mentioned already, TD-EDF cannot spontaneously lead to 
the separation of the system into two fragments below $Q^{\rm th}_2$ due to the improper treatment of level crossing.
Still it is possible to induce a fission by imposing some collective velocity initially. As already noted 
in Ref. \cite{God14}, the collective energy that should be initially put in the system to observe fission 
is in general rather large. This aspect is further illustrated in Fig. \ref{fig:eth} where we investigated 
systematically the minimal collective energy necessary to induce fission for selected initial quadrupole 
deformation. To obtain this curve, for each $Q^{\rm ini}_2$ we systematically performed TD-EDF calculation by increasing 
progressively the boost intensity. The error bars correspond respectively to the largest (resp. lowest) collective energy 
where fission is (resp. is not) observed. Note that for the two lowest $Q^{\rm ini}_2$ no binary fission but ternary fission 
is observed (see also Fig. \ref{fig:denstime}). 

The collective energy are very high compared to the typical barrier height to fission.
We would like to mention that this is clearly a pathology of TD-EDF at small initial deformation 
and beyond mean-field effects should clearly be included to obtain meaningful 
information from microscopic transport models around the fission barrier. Still, to illustrate that 
the present method can apply in situation rather far from the adiabatic limit, we deduced the mass parameter 
for such initially highly excited systems. A few examples are shown in panel (b) of Fig. \ref{fig:mass_b14}. In that case, 
important deviation are observed in the mass parameters compared to the static/adiabatic limit. Since large collective 
velocities are imposed initially, such deviations are not surprising. However,  the difference can also stem from the fact that the 
initial boost can induce a motion that is not described by the simple one-dimensional energy landscape shown in Fig. \ref{fig:pes}. In particular, 
preparing the system using constraint mean-field + boost is a rather arbitrary choice that will induce specific motion not only 
in the $Q_2$ collective space but also in a larger space of collective variables like the monopole  $\langle \bvec r^2 \rangle$, 
hexadecapole $Q_4$, .... In particular, since the mass reported in Fig. \ref{fig:mass_b14} are compared for the 
same $Q_2$, differences observed between the static and dynamical masses stem from differences in the root mean-square radii 
that ultimately come from the differences in local densities. Clear differences are observed between densities shown in Fig. 
\ref{fig:denstime} and densities of the adiabatic PEC (Fig. \ref{fig:pes}). The differences in local densities can of course 
come from non-adiabatic effects but also from a more complex path in a multi-dimensional potential energy 
landscape that could not be simply reduce to the 1D picture of Fig. \ref{fig:pes}.

\subsection{Total versus collective kinetic energy}

As we mentioned in Sec. \ref{sec:mass}, 
the present method allows us to access the set of conjugate momenta and 
the collective kinetic energy (CKE) as well as the masses 
associated with the set of collective coordinates. 
The CKE of the set of collective variables $\{Q_\alpha\}$ can be obtained following 
section \ref{sec:diagm} diagonalizing the mass matrix and using  Eq. (\ref{eq:ekingen}). Note that the diagonal and off-diagonal matrix 
elements of the mass are given explicitly in appendix \ref{sec:appendix}.

In Fig. \ref{fig:tcoll}, the CKE associated to the quadrupole and/or hexadecapole moment are displayed as a function 
of time during the fission process. We also compare these energies to  the total kinetic energy computed  through
\begin{eqnarray}
E^{\rm tot}_{\rm kin}=\frac{\hbar^2}{2m}\int d^3r\ \frac{\bvec j(\bvec r, t)^2}{n(\bvec r, t)}, 
\label{eq:ektot}
\end{eqnarray}
where $\bvec j(\bvec r , t)$ is the single-particle current. Two different initial conditions are considered, one starting from an 
already elongated  shape without boost and one with a more compact shape but where a boost in quadrupole momentum is 
applied to induce fission.

\begin{figure}
\begin{center}
\includegraphics[scale=0.9]{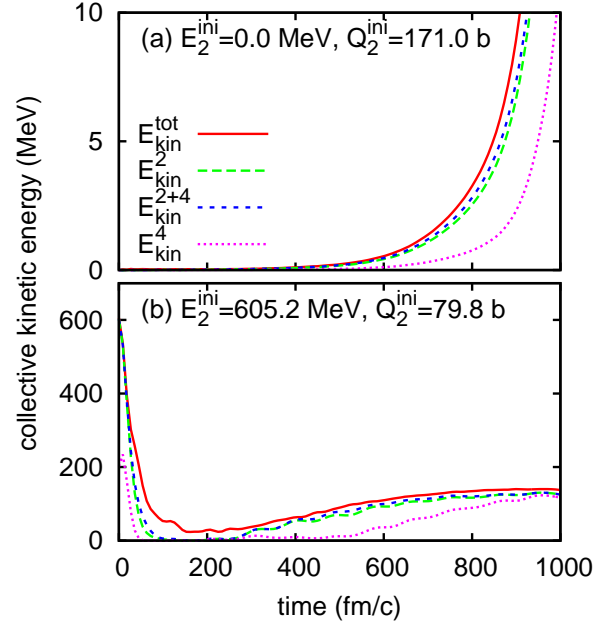}
\end{center}
\caption{(Color online) Evolution of the total collective kinetic energy $E^{\rm tot}_{\rm kin}$ as a function of time. 
The initial systems correspond 
respectively to a quadrupole moment (a) $Q^{\rm ini}_2=171.0$ b without boost or (b) to $Q^{\rm ini}_2=79.8$ b with initial boost. 
In both cases, the CKE obtained using Eq. (\ref{eq:ekingen}) and associated to  $Q_2$ only ($E^{2}_{\rm kin}$), $Q_4$ only ($E^{4}_{\rm kin}$) or both  ($E^{2+4}_{\rm kin}$) 
are also shown.}
\label{fig:tcoll}
\end{figure}

From this figure several interesting aspects could be seen:
\begin{itemize}
  \item At initial time $E^{\rm tot}_{\rm kin} = {E}^{2 }_{\rm kin}$. This is indeed due to the fact that either the two are equal to zero (Fig. \ref{fig:tcoll}(a))
  or that the initial condition (Fig. \ref{fig:tcoll}-b) is such that all initial kinetic energy is imposed by the quadrupole boost.  
  \item The CKE associated to $Q_4$ is also initially non zero. This stems from the fact that $Q_2$ and $Q_4$ are not independent 
  collective variables. Therefore boosting in quadrupole moment also induces an excitation of the hexadecapole and most probably  
  higher order even multipole moments.
  \item Due to the rather strong correlations between $Q_2$ and $Q_4$, the off diagonal matrix elements of the inertia play 
  an important role. Indeed, neglecting this contribution would give:
  \begin{eqnarray}
E^{2+4}_{\rm kin} \simeq E^{2}_{\rm kin} + E^{4}_{\rm kin}.
\end{eqnarray} 
However, summing directly these two energies would exceed the total kinetic energy that is an upper bound whatever is the selected 
set of collective variables. In Fig. \ref{fig:tcoll},  $E^{2+4}_{\rm kin}$ accounts  for the off-diagonal inertia and finally leads to an energy 
that is lower than $E^{\rm tot}_{\rm kin}$. 
\item At large distances, we see that 
\begin{eqnarray}
E^{\rm tot}_{\rm kin} \simeq E^{2+4}_{\rm kin} \simeq E^{2}_{\rm kin}. 
\end{eqnarray}
This is due to the fact that all kinetic energies are dominated by the relative motion of the two fragments in the exit channel.  
\end{itemize}
\begin{figure}
\begin{center}
\includegraphics[scale=0.7]{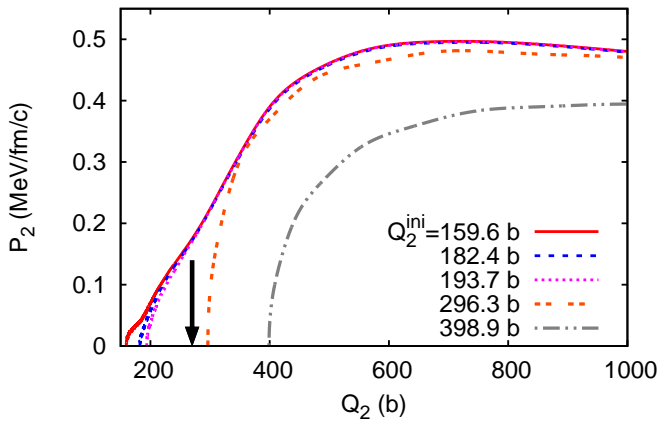}
\end{center}
\caption{(Color online) Evolution of the collective momentum as a function of time for different 
initial quadrupole deformations (with $Q^{\rm ini}_2 \ge Q^{\rm th}_2 $). The arrow indicates the scission point 
associated to the adiabatic potential. }
\label{fig:P2dP2}
\end{figure}

\subsection{Collective evolution close to scission} 

Here we investigate the collective evolution close to the scission point. The scission 
can be seen directly on Fig. \ref{fig:pes} by the change of slope around $Q^{\rm sc}_2 \simeq 270$ b.
The evolution of the collective momentum is displayed in Fig. \ref{fig:P2dP2} as a function of $Q_2$
for different initial deformations. 
We clearly see a different behavior depending if the initial quadrupole moment is above or below 
$Q^{\rm sc}_2$. As we will see below, for $Q^{\rm ini}_2 \ge Q^{\rm sc}_2$ the momentum evolution 
corresponds essentially to the evolution of two escaping nuclei boosted by their mutual Coulomb field.    
For $Q^{\rm ini}_2 \le  Q^{\rm sc}_2$, the nuclear interaction between nuclei still plays a significant role and a richer 
evolution is seen.  In that case, independently of the initial $Q^{\rm ini}_2$ value, after some transition time, all curves 
become nearly identical with one another.  

In the absence of dissipation and assuming that the dynamics stem uniquely from a collective potential, one would expect that 
the smaller is $Q^{\rm ini}_2$, the higher is $P_2(t)$ as a function of $Q_2$. However, it clearly seems from Fig. \ref{fig:P2dP2} that part of the energy 
is dissipated in the early stage of the evolution.  To further progress, we may follow Ref. \cite{Was08} and assume that the 
momentum evolution can be written as a simple dissipative equation of motion:
\begin{eqnarray}
\dot P_2 = -\frac{\partial V_{coll}}{\partial Q_2}
+\frac{1}{2}\frac{\partial M_2}{\partial Q_2}\dot Q_2^2
-\gamma(Q_2)\dot Q_2, \label{eq:colldiss}
\end{eqnarray}
where the collective potential $ V_{coll}$ and the friction coefficient $\gamma$ 
are a priori unknown quantities. In the adiabatic limit, the collective potential 
identifies with the one shown in Fig. \ref{fig:pes}  and $\gamma(Q_2) = 0$ along the path.

\begin{figure}
\begin{center}
\includegraphics[scale=1.15]{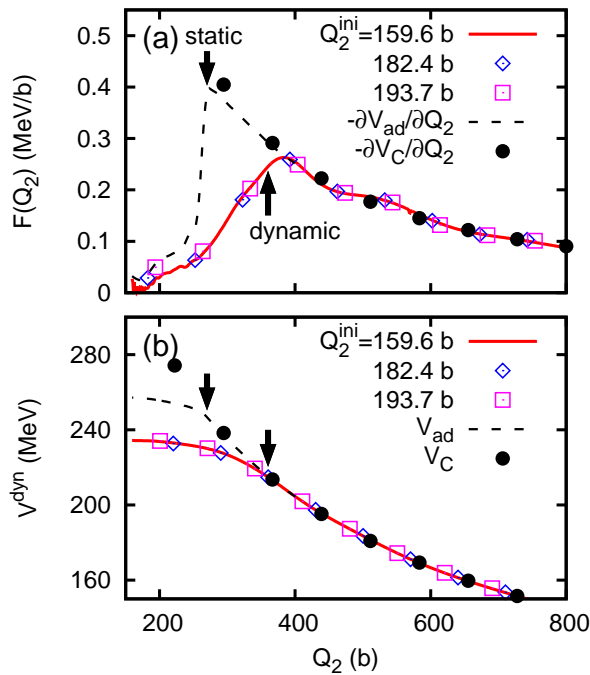}
\end{center}
\caption{(Color online) (a) Function $F(Q_2)$ obtained with TD-EDF using Eq. (\ref{eq:force})
for the three evolutions with $Q^{\rm ini}_2 \leq Q^{\rm sc}_2$ displayed in Fig. \ref{fig:P2dP2}.
For comparison, we also show the forces acting on $Q_2$ that would be induced either by
the adiabatic potential (black dashed line) or solely by the Coulomb field (black filled circle).
The arrows in the figure indicate the $Q_2$ value where the neck density $\rho_{\rm neck}$ 
becomes ten times less than the saturation density $\rho_{\rm sat}=0.16$ fm$^{-3}$. 
The two arrows indicates the adiabatic path (static) and dynamical path (dynamic). In the latter case, 
the position where $\rho_{\rm neck}/\rho_{\rm sat} = 10$ is almost independent of $Q_2^{\rm ini}$. (b) Dynamical potential curve obtained 
by integrating $F(Q_2)$ using Eq. (\ref{eq:vdyn}). 
Again for comparison, the adiabatic potential and the Coulomb field are also shown. 
Here, the origin of energy is taken such that $E=0$ for $Q_2\to\infty$. }
\label{fig:force}
\end{figure}

To remove the possible effect of the mass evolution and eventually access the potential and dissipative collective properties, 
it is convenient to define the quantity
\begin{eqnarray}
F(Q_2)\equiv \dot P_2-\frac{1}{2}\frac{\partial M_2}{\partial Q_2}\dot Q_2^2. 
\label{eq:force}
\end{eqnarray}
This function is shown in Fig. \ref{fig:force} as a function of $Q_2$ for some of the evolutions 
presented in Fig. \ref{fig:P2dP2}. If the macroscopic transport equation (\ref{eq:colldiss}) is valid, this quantity 
is expected to identify with:
\begin{eqnarray}
F(Q_2) = -\frac{\partial V_{coll}}{\partial Q_2}-\gamma(Q_2)\dot Q_2.
\label{eq:force2}
\end{eqnarray}
and therefore is sensitive to both the potential and dissipative part.
For comparison, we also display the cases where dissipation is assumed 
to be zero in Eq. (\ref{eq:force2}) and where the potential part identifies 
either to the adiabatic potential or solely to the Coulomb field. In the latter 
case, the Coulomb potential for large relative distance, or large $Q_2$, is approximated by 
\begin{eqnarray}
V_C\approx\frac{Z_1Z_2e^2}{R}
\approx \frac{Z_1Z_2e^2}{\sqrt{\frac{A}{2A_1A_2}Q_2}}
= \frac{1}{4}\sqrt{\frac{A}{2}}Z^2e^2Q_2^{-1/2}.
\end{eqnarray}
The last expression is obtained for the symmetric fission case, i.e. $Z_1=Z_2 = Z/2$ and  
$A_1=A_2 = A/2$ considered here, and further assuming no intrinsic quadrupole deformation of 
emitted fragments after scission. 

Fig. \ref{fig:force}-a gives interesting information on the different steps leading to fission. We first see that after the very first 
instant of the evolution where some dissipation occurred, all evolutions obtained with $Q^{\rm ini}_2 \le Q^{\rm sc}_2$   
are on top of each other. The dynamic before scission deviates significantly from the expected adiabatic one underlining 
the importance of both non-adiabatic and dissipative effects.  In particular, we clearly see that the dynamical 
formation of the neck differs from the adiabatic case. Defining the scission point as the $Q_2$ value  where 
the neck density equal $1/10$ of the saturation density, we observe that dynamically the scission occurs at much larger $Q_2$
than the adiabatic case (arrows in Fig. \ref{fig:force}). This has two consequences (i) first, the two nuclei stick together at larger distance compared 
to the adiabatic case. Accordingly, the nuclear field can play an enhanced role. (ii) we see that we should introduce the notion of ``dynamical scission point''  
that a priori differs from the ``adiabatic scission point'' and that occurs at larger quadrupole moment. In the present case of symmetric compact fission, 
the dynamical scission point occurs around $Q^{\rm sc,dyn}_2 \simeq 360$ b, compared to the adiabatic scission point $Q^{\rm sc,stat}_2 \simeq 270$ b. 

After scission, the dynamical evolution is very close to the Coulomb field case (black filled circles). 
This indicates that no dissipation takes place after this point and that the dissipation mainly occurs 
at initial time of the calculation. For large $Q_2$,
we clearly observe some oscillations around the average Coulomb repulsion 
that could be attributed to the dynamical oscillation of the intrinsic shapes 
of each nucleus. 
These oscillations obviously goes beyond the simple macroscopic approximation (\ref{eq:colldiss}) since they involves additional intrinsic 
shape degrees of freedom. 

Following \cite{Was08}, one could a priori use $F(Q_2)$ to get the potential energy landscape as well as 
the friction coefficient along the path. However, the method used in Ref.  \cite{Was08} that consists in 
performing two evolutions with close initial conditions cannot be applied here due to the fact that the collective velocity
becomes rapidly independent of $Q^{\rm ini}_2$. Fig. \ref{fig:P2dP2} seems however to indicate that dissipation 
occurs only at rather small $Q_2$. For $Q_2> 300$ b, one might assume that the motion is only driven by a potential denoted 
by $V^{\rm dyn}(Q_2)$. Then, we have the approximate relationship:
\begin{eqnarray}
\frac{\partial V^{\rm dyn}(Q_2) }{\partial Q_2} &=& - F(Q_2)
\end{eqnarray}     
where $ F(Q_2)$ is estimated along the path using Eq. (\ref{eq:force}). To get the potential itself, one should fix the boundary condition. 
We know from Fig. \ref{fig:force}-a that the potential identifies to a good approximation with the Coulomb potential after the scission. The potential
can eventually be obtained through the relation:
\begin{eqnarray}
V^{\rm dyn} (Q_2) & = & V_C (Q^{\rm max}_2)  + \int^{Q^{\rm max}_2}_{Q_2} F(Q_2') dQ_2' \label{eq:vdyn}
\end{eqnarray}  
where $Q^{\rm max}_2$ is taken much larger than the dynamical scission point. 
{Note that we do not take into account here the excitation energy of the fragments. } 
Examples of potential obtained in this way is shown in Fig. \ref{fig:force}-b 
assuming $Q^{\rm max}_2 = 800$ b. 
$V_{\rm ad}$ shown with the dashed curve in Fig. \ref{fig:force}-b is drawn by shifting the PEC given in Fig. \ref{fig:pes} 
so that it coincides with $V_C$ at $Q_2=433$ b ($\geq Q_2^{\rm sc,stat}$) and thus $V_{\rm ad}\to 0$ for $Q_2\to\infty$. 

We see in Fig. \ref{fig:force}-b that the potential obtained using Eq. (\ref{eq:vdyn}) differs significantly from the adiabatic 
one at small $Q_2$ due to dynamics and eventually non-adiabatic effects. Note that the dynamical potential should 
be interpreted with some caution since it might contain some dissipative effects especially at initial time.
It is worth in particular mentioning that the adiabatic and dynamical potentials should be identical at initial $Q_2$. 
We clearly observe in Fig. \ref{fig:force}-b a lower value for the dynamical case. 
The difference between the adiabatic 
and dynamical curves at $t=0$ corresponds to the energy transferred into the other collective 
degrees freedom or internal excitations during the fission. 
We see that this difference is $\approx 23$ MeV for $Q_2^{\rm ini}=160$ b. 

\subsection{Dissipation estimated from energy balance}

Alternatively, in order to estimate the energy dissipated into the internal excitation of the fragments, 
an analysis similar to Ref. \cite{Sim14} 
has been made for the total kinetic energy (TKE) of the outgoing fragments after fission. 
The TKE is defined as 
\begin{eqnarray}
{\rm TKE}= \frac{1}{2}\mu\dot R^2+\frac{Z_1Z_2e^2}{R}
 \equiv T_{\rm rel}+V_C(R)\end{eqnarray}
at large $R$ well beyond the scission point. 
Note that the TKE is the energy of the relative motion between the fragments at infinite 
separation, and it is not the same quantity as either Eq. (\ref{eq:ekingen}) or Eq. (\ref{eq:ektot}), 
which may contain the internal excitation energy of the fragments. 
The energy dissipated into the other DOFs than the relative motion, or the 
excitation energy of fragments, is then given by $E^*=E_0-{\rm TKE}$, where $E_0$ is the 
total energy of the system \cite{Sim14}. Note that the origin of 
energy for $E_0$ is taken as that for $R(t) \to\infty$ with fragments staying at their ground states. 

\begin{figure}
\begin{center}
\includegraphics[scale=.6]{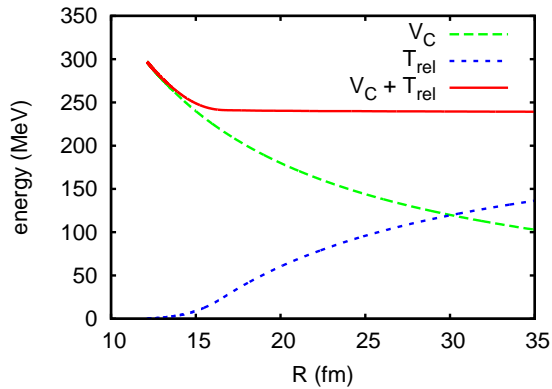}
\end{center}
\caption{The relative kinetic energy $T_{\rm rel}$, the Coulomb energy $V_C$, 
and the total kinetic energy (TKE) 
of the fission fragments as a function of the relative distance $R$ for $Q_2^{\rm ini}=197$ b. }
\label{fig:TKE}
\end{figure}

In Fig. \ref{fig:TKE} we show relative kinetic energy $T_{\rm rel}$, the Coulomb energy $V_C$, 
and the total kinetic energy (TKE) of the fission fragments as a function of the relative distance 
$R$ for the evolution with $Q_2^{\rm ini}=194$ b. 
We see a plateau in TKE for large $R$ at $\approx 238$ MeV, which is 
identified as the TKE for this process, while the total energy of the system is $E_0\approx 256$ MeV, which 
is given by the value of $V_{\rm ad}$ in Fig. \ref{fig:force}-b at $Q_{2}^{\rm ini}$. 
Taking the difference, we obtain $E^*\approx 18$ MeV. This is close to the value obtained
from the previous analysis confirming that the difference between the adiabatic potential and the dynamical 
potential estimated from Eq. (\ref{eq:vdyn}) most likely stems from the energy dissipated along the fission path. 

\section{Conclusion}

In the present work, a method is proposed to construct conjugated collective momenta
associated to a given  set of local collective variables along a time-dependent EDF path.  
A detailed discussion is made on
the proper definition of associated inertia including the effect of its possible off-diagonal   
matrix elements. Once pairs of conjugated collective  variables are obtained, one can 
make a macroscopic reduction of the microscopic mean-field dynamic. 

An illustration
is given here  with the fission process. A precise analysis is made in the symmetric fission
of $^{258}$Fm. The mass matrix is calculated along the fission path including only quadrupole moment, and/or both
the quadrupole and hexadecapole moment. In particular, the important role of off-diagonal matrix elements of
the mass is underlined. 
Then, a detailed analysis of the macroscopic evolution in the quadrupole collective space is made. The importance 
of dissipation in the early stage of the evolution is discussed. Clear non-adiabatic effects are probed, in particular
associated with the specific neck evolution. We show that the scission point, called dynamical scission point, occurs 
at much larger quadrupole moment compared to that for the adiabatic path. 
An attempt to extract the nucleus-nucleus 
potential felt by the daughter nuclei after fission is also made. 
It is shown that this potential significantly differs from the adiabatic one due to the non-adiabatic effects 
and the dissipation of energy into the intrinsic excitations of fission fragments.  

The method presented here is rather versatile and could be used in other dynamical processes. 
For instance, it could a priori be used to study anharmonicity in giant resonances as well as 
possible coupling between collective modes.

\begin{acknowledgments}
We would like to thank G. Adamian, N. Antonenko, S. Ayik and K. Hagino for useful discussions
on collective observables.  
G.S. acknowledges the Japan Society for the Promotion of Science
 for the JSPS postdoctoral fellowship for foreign researchers.
 This work was supported by Grant-in-Aid for JSPS Fellows No. 14F04769. 

\end{acknowledgments}

\appendix
\section{General formula for the inertia matrix}\label{sec:appendix}

In this appendix we give the explicit expression for the inertia matrix 
\begin{eqnarray}
\frac{m}{M_{\alpha\beta}}={\rm Tr}[\rho \nabla Q_\alpha\cdot\nabla Q_\beta], 
\end{eqnarray}
associated with the general multipole moment 
\begin{eqnarray}
\hat Q_{\lambda}=\sqrt{\frac{16\pi}{2\lambda+1}}r^\lambda Y_{\lambda 0}. 
\end{eqnarray}
Its expressions for $2\leq\lambda\leq 6$ are given in Table \ref{tab:multipole}. 

\begin{table}
\caption{Multipole operators $\hat Q_\lambda=\sqrt{\frac{16\pi}{2\lambda+1}}r^\lambda Y_{\lambda 0}$ 
for $\lambda=2,3,4,5,6$. }
\begin{center}
\begin{tabular}{c|l}
\hline\hline
$~~\lambda~~$ & $~~$operator \\
 \hline
2 & $~2z^2-x^2-y^2$ \\ [.2cm]
3 & $~2z^3-3zx^2-3y^2z$ \\[.2cm]
4 & $~\displaystyle{\frac{1}{4}(35z^4-30r^2z^2+3r^4)}$ \\ [.2cm]
5 & $~\displaystyle{\frac{1}{4}(63z^5-70r^2z^3+15r^4z)}$ \\ [.2cm]
6 & $~\displaystyle{\frac{1}{8}(231z^6-315r^2z^4+105r^4z^2-5r^6)}$ \\ [.2cm] 
\hline\hline
\end{tabular}
\end{center}
\label{tab:multipole}
\end{table}

Using the   Racah algebra technique, one obtains the following for the matrix 
\begin{widetext}
\begin{eqnarray}
\frac{m}{M_{\lambda\lambda'}}=\sum_{L}(2L+1)
\frac{(s-2L)!(s-2\lambda)!(s-2\lambda')!(s/2)!(s/2-1)!}
{(s-1)!(s/2-L)!(s/2-L-1)![(s/2-\lambda)!]^2[(s/2-\lambda')!]^2}
\cdot {\rm Tr}[\rho r^{\lambda+\lambda'-2-L}\hat Q_L], 
\end{eqnarray}
where $s=\lambda+\lambda'+L$ and the value of $L$ runs over 
$L=|\lambda-\lambda'|, |\lambda-\lambda'|+2, |\lambda-\lambda'|+4, \ldots, \lambda+\lambda'-2$. 
It allows us to compute a matrix element for any multipolarities $\lambda$ and $\lambda'$. 
In Table \ref{tab:massmat} we illustrate expressions of the matrix elements for 
some important multipoles 2, 3, and 4, which are be related 
to elongation, mass asymmetry, and size of neck, respectively.

\begin{table}
\caption{Illustration of the inertia matrix elements $m/M_{\lambda\lambda'}$ for the multipoles 
of $\lambda,\lambda'=2,3,4$. The expressions for the multipole operators are given in 
Table \ref{tab:multipole}. }
\begin{center}
\begin{tabular}{c|ccc}
\hline\hline
$\lambda~\backslash~\lambda'$ & 2 & 3 & 4 \\
 \hline
2 & $4\bigl(2\langle r^2 \rangle+\langle Q_2\rangle\bigr)$ 
  & $\displaystyle{\frac{12}{5}\bigl(3\langle r^2Q_1 \rangle + 2\langle Q_3\rangle\bigr)}$ 
  & $\displaystyle{\frac{8}{7}\bigl(9\langle r^2Q_2 \rangle + 5\langle Q_4\rangle\bigr)}$ \\ [.2cm]
3 & --- 
  & $\displaystyle{\frac{12}{7}\bigl(7\langle r^4 \rangle + 4\langle r^2Q_2\rangle+3 \langle Q_4\rangle\bigr)}$ 
  & $\displaystyle{\frac{8}{77}\bigl(99\langle r^4Q_1 \rangle + 77\langle r^2Q_3\rangle+ 150\langle Q_5\rangle\bigr)}$\\ [.2cm]
4 & ---
  & ---
  & $\displaystyle{\frac{8}{231}\bigl(462\langle r^6 \rangle + 275\langle r^4Q_2\rangle+ 243\langle r^2Q_4\rangle+175\langle Q_6\rangle\bigr)}$ \\ [.2cm] 
\hline\hline
\end{tabular}
\end{center}
\label{tab:massmat}
\end{table}

\end{widetext}

\section{Definition of new operators $(Q'_\alpha, P'_\alpha)$}\label{sec:mdiag}
In this appendix, we give some intermediate steps to obtain operators that fulfill the commutation 
rules (\ref{eq:conjstrong}) along the TD-EDF trajectory.

The momentum operator conjugated to the coordinate $Q_\alpha$ is given by 
\begin{eqnarray}
P_\alpha = -i\hbar\frac{M_{\alpha\alpha}}{m}\left(\frac{\nabla^2Q_\alpha}{2}
+\nabla Q_\alpha \cdot \nabla \right). 
\end{eqnarray}
Accordingly, we have:
\begin{eqnarray}
\langle[Q_\alpha,P_\beta]\rangle&=&
i\hbar\frac{M_{\beta\beta}}{m}\langle \nabla Q_\alpha\cdot\nabla Q_\beta\rangle . \nonumber
\end{eqnarray}
Introducing the mass matrix elements $m/M_{\alpha\beta}\equiv \langle \nabla Q_\alpha\cdot\nabla Q_\beta\rangle$, above expression 
can be written as
\begin{eqnarray}
\langle[Q_\alpha,P_\beta/M_{\beta\beta}]\rangle = i\hbar\frac{1}{M_{\alpha\beta}}.
\end{eqnarray}
We introduce the orthogonal matrix $W$ that diagonalizes the mass, i.e. 
\begin{eqnarray}
\sum_{\alpha\beta}W_{k\alpha}\frac{1}{M_{\alpha\beta}}W^T_{\beta l} = \delta_{kl}\frac{1}{M_k'}
\end{eqnarray}
where $1/M'_k$ are the eigenvalue of the inertia tensor. 
Then we have 
\begin{eqnarray}
\sum_{\alpha\beta}W_{k\alpha}W^{T}_{\beta l}\langle[Q_\alpha,P_\beta/M_{\beta\beta}]\rangle
=i\hbar\delta_{kl}\frac{1}{M_{k}}. 
\end{eqnarray}
Introducing the new set of conjugated operators 
\begin{eqnarray}
Q_k'=\sum_\alpha W_{k\alpha}Q_\alpha,~{\rm and}~
P_k'=M_k'\sum_\alpha W_{k\alpha}\frac{P_\alpha}{M_{\alpha\alpha}},
\end{eqnarray}
we see that these operators respect the desired commutation relation 
\begin{eqnarray}
\langle[Q_k',P_{l}'/M_l']\rangle= i\hbar\delta_{kl}\frac{1}{M_{k}'}
\end{eqnarray}
or equivalently
\begin{eqnarray}
\langle[Q_k',P_l']\rangle = i\hbar\delta_{kl}. 
\end{eqnarray}

\end{document}